# Short-term Volatility Estimation for High Frequency Trades using Gaussian processes (GPs)

Leonard Mushunje[1*], Maxwell Mashasha[2] and Edina Chandiwana[3]

Midlands State University,

Department of Applied Mathematics and Statistics

leonsmushunje@gmail.com

**Abstract**

The fundamental theorem behind financial markets is that stock prices are intrinsically complex and stochastic. One of the complexities is the volatility associated with stock prices. Volatility is a tendency for prices to change unexpectedly [1]. Price volatility is often detrimental to the return economics, and thus, investors should factor it in whenever making investment decisions, choices, and temporal or permanent moves. It is, therefore, crucial to make necessary and regular short and long-term stock price volatility forecasts for the safety and economics of investors' returns. These forecasts should be accurate and not misleading. Different models and methods, such as ARCH GARCH models, have been intuitively implemented to make such forecasts. However, such traditional means fail to capture the short-term volatility forecasts effectively. This paper, therefore, investigates and implements a combination of numeric and probabilistic models for short-term volatility and return forecasting for high-frequency trades. The essence is that one-day-ahead volatility forecasts were made with Gaussian Processes (GPs) applied to the outputs of a Numerical market prediction (NMP) model. Firstly, the stock price data from NMP was corrected by a GP. Since it is not easy to set price limits in a market due to its free nature and randomness, a Censored GP was used to model the relationship between the corrected stock prices and returns. Forecasting errors were evaluated using the implied and estimated data.

## 1. Introduction

Stock prices are geared towards determining investors' portfolio status, and their consideration is essential not only in stock markets. In South Africa, Stocks listed at the Johannesburg Stock Exchange (JSE) are going through volatilities whose coefficients are high, and this is most common in almost all emerging economies. The frequency of the short-scaled volatility hits often poses several investment and operational challenges. Thus, volatility prediction is essential for securing the economies of investors' investment portfolios. Short-term volatility forecasts with a prediction horizon from one hour to several days are critical to optimize stock returns and any associated costs.

There are two approaches to short-term price volatility forecasting: statistical models and physical models. The former uses only historical stock price data to build statistical models, such as autoregressive integrated moving average (ARIMA), autoregressive conditional heteroscedasticity (ARCH), generalized autoregressive conditional heteroscedasticity (GARCH), artificial neural networks, Kalman filters, support vector machines. The cross-field application of these models appears in wind power generation forecasts (see [2-6]). Since prices vary rapidly with time, the statistical models are effective only for very short-term forecasts (about 1–3 hours ahead). On the other hand, physical models have advantages over longer horizons (days, weeks, months) because they include (3-Dimension) spatial and temporal factors in a full fluid-dynamics model. However, this model type has limitations, such as the limited observation set for model calibration. To overcome these limitations, some authors have combined statistical and physical models [7,8], where data from a physical model is used as inputs to a statistical model.

This study proposes a forecast model combined with NMP data in which one-day-ahead price volatility forecasting is realized based on historically recorded close prices, volumes, and other market information. We shall combine our NMP data with the Gaussian Processes (GP). Such integrated methods will be used to make corresponding return predictions. Related to our study are the works of [9] who examines the accuracy of several of the most popular methods used in volatility forecasting. A comparative approach is employed where historical volatility models such as the Exponential Weighted Moving Average, ARMA model, and GARCH family of models are compared with Artificial Neural Networks-based models. [10] proposed a simple but less accurate method of estimating volatility where daily squared returns are taken. The jumps associated with intra-day prices are not captured, yet these jumps significantly affect volatility.

Other related works were done on the prediction of the stock/index returns by [11-15], where Artificial neural networks were used. In addition, [16] employs a variation of a type of Recurrent Neural Network called Long-Short Term Memory (LSTM) to predict stock price volatility in the US equity market. Among their results, they found that a more incredible deal of tuning is required on the deeper network, and in particular, the increased use of dropout layers could help reduce the variance problem associated with the employed model to estimate the price volatilities accurately. Recent work on stock market prediction is by [17] who focus on applying LSTM to predict financial time series in the stock market, using both traditional time series analysis and technical analysis metrics. This is directly related to the successful application of Long Short-Term Memory (LSTM) to address the problem of volatility prediction in the stock market, [18-20]. On the other hand, [21] provides a literature review using a systematic database to examine and cross-reference snowballing where previous studies featuring a generalized autoregressive conditional heteroskedastic (GARCH) family-based model stock market return and volatility are reviewed. They also conducted a content analysis of return and volatility literature reviews for 12 years (2008–2019) and in 50 different papers to see the trends and concentration of volatility-linked studies. Concerning volatility and deviation modeling, researchers have proposed other distributed models to describe better the thick tail of the daily rate of return. For instance, [22] first proposed an autoregressive conditional heteroscedasticity model (ARCH model) to characterize some possible correlations of the conditional variance of the prediction error. In 1986, Bollerslev extended the ARCH model to form a generalized autoregressive conditional heteroskedastic (GARCH) model. Later, the GARCH model rapidly expanded to other forms, such as TARCH, EGARCH, and ETARCH, to form the GARCH family. As indicated across the literature, researchers proved that GARCH is the most suitable model to use when one has to analyze the volatility of the returns of stocks with enormous volumes of observations (for more, see [22]; [23-26].

From the reviewed literature, short-term volatility forecasting has been slimly done, and more attention needs to be paid to jumps in association with these short-timed price swerves. Statistical models have been employed for the volatility studies, as stated earlier in this paper. In this paper, stock prices and related factors such as returns and volumes' datasets, including Numerical market prediction (NMP) results, are analyzed and used to develop volatility forecasting models over a horizon of up to one day, with a Gaussian Process (GP) method. This paper's main contributions and thrust can be summarized into four categories: 1. The predicted price volatility from an NMP model is corrected using a GP. This process helps to improve performance compared with earlier methods for combining statistical and physical models. 2. A censored Gaussian Process (CGP) method is applied to build the relationship between corrected stock prices and stock volumes. The method accounts for the probabilistic character of the values that are not known precisely because of censoring. 3. A subset of high-stock price data is treated separately because of its different characteristics based on analysis of the initial models. 4. Historical stock price data from the JSE databases is used as an additional input to the forecasting model for 1–3 hours-ahead prediction since we have proved this effective in this range of time horizons. The idea paves an excellent way for high-frequency trades that are proving to dominate the markets and investment world.

## 2. Methodology

### 2.1. Data

The datasets used in this study were extracted from the Johannesburg Stock Exchange (JSE) recent stock price databases from January 2010 to January 2023. We shall use a whole year dataset as a training set and the remainder as an independent test set, from where we will make our suitable inferential conclusions. The missing values were less than 30%, and to cater to them, we used the K-nearest neighbor (KNN).

### 2.2. Numerical market prediction Model and Volatility forecasting

Numerical market prediction uses statistical physics and statistical historical models related to financial markets' mechanics. They predict prices based on specific initial values and boundary conditions. This study uses the stock price data (JSE) and the SPD-NVP model. SPD (stock price data) is extracted from the frequently updated JSE electronic stock price databases. The databases are prepared and well-kept for the interests of investors. In general, short-term price volatility forecasting needs predictions from an NMP model with high spatial resolution. The stock price data from JSE is suitable for this application. Hence, there is no need for extra actions like backward and forward interpolation. The prediction data is produced daily and is usually available at 4:00 PM CAT when closing valuations are done for most investment assets. The data, including stock prices, volumes, and returns, is provided for 30 minutes for the following 24 hours. It is no secret that investors and market regulators require accurate stock price and return forecasts. In this study, we stress that the forecasting error of 1– 3 hours ahead should be less than 10% of the actual recorded figures. Therefore, all the forecast errors contained in this study are calculated using hourly data.

## 3. The Gaussian Process (GP)

The method of Gaussian processes has been introduced previously. However, less is considerably known about its application to financial data. Moreover, it has been successfully applied to many machine-learning tasks. [27] duped a well-detailed systematic explanation of Gaussian process regression and Automatic Relevance Determination (ARD). The reader is encouraged to consult this book for more information on the GPs. Further, the Gaussian Processes (GP) extension to censored data is found in [28]. However, this study only provides a brief description.

### 3.1. Gaussian Process model

Let us consider a Gaussian process $f(x)$ for a classic regression problem. Now, assuming that we have training set $D$ with $n$ observations such that

$D = \{(x_i, y_i) | i = 1, \ldots, n\}$, where $x$ denotes an input vector, and $y$ denotes a scalar output. The task is to build a function that satisfies the following multiple linear equation.

$$\boldsymbol{y_i} = \boldsymbol{f(x_i)} + \boldsymbol{\epsilon_i},,,,,,,,,,,,,,(\mathbf{1})$$

$\boldsymbol{\epsilon_i}$ **is** the non-observable additive noise parameter and is assumed to follow a Gaussian distribution such *as* $\boldsymbol{\epsilon_i} \sim N(0, \sigma_n^2)$. Note that $y$ is a linear combination of Gaussian variables; hence, using the invariant transformation property of linear functions is itself Gaussian. Therefore, we have

$p(y|X, k) = N(0, K + \sigma_n^2 I)$ , where $K_{ij} = k(x_i, x_j)$, and the joint distribution for a new input $x_*$ can be written in matrix form as:

$$\begin{bmatrix} y \\ f_* \end{bmatrix} \sim \left(0, \begin{bmatrix} K(X,X) + \sigma_n^2 & k(X, x_*) \\ k(x_*, X) & k(x_*, x_*) \end{bmatrix}\right),,,,,,,,,,(\mathbf{2}),$$

where, $k(X, x_*) = k(x_*, X)^T = [k(x_1, x_*), \ldots, k(x_n, x_*)]$, which we will shortly express as $k_*$. Consequently, following the properties of joint Gaussian distributions, we predict the distribution of our target variable using the following function:

$$\overline{f_*} = k_*^T (K + \sigma_n^2 I)^{-1} y,,,,,,,,,,,,,,,,,,,,,,,,,,,(\mathbf{3})$$

$$V[f_*] = k(x_*, x_*) - k_*^T (K + \sigma_n^2 I)^{-1} k_*$$

As a result of the stock price control strategies available in the market, there is always a defined upper limit $\boldsymbol{S_{upper}}$ and lower limit of 0 for the stock prices placed at JSE. Therefore, in statistics, the actual values (unrestricted price output) are 'censored' in that they are not observed but are replaced by the threshold value. Our analysis assumes that the latent values $y^* = f(x)$ can be realized by a Gaussian process. Thus, to predict the actual output$(y)$, the influence of censoring is considered inside the model. We used the censored GP model [28] developed to implement this consideration. In the model, the posterior distribution of $y$ is formed by integrating the censored distribution of latent variables and approximated using Expectation Propagation (EP). Exploratory data analysis shows that 4.6% percent is within 5% of the upper limit. Thus, it is in the range where the noise distribution overlaps significantly with the censored range. Therefore, it is essential to account for this constraint in the model itself rather than simply pre-processing the predictions of a 'standard' regression model by thresholding them at$\boldsymbol{S}_{\boldsymbol{upper}}$.

## 4. Modelling Process

Our modeling process follows the same approach used in wind power prediction [29] demonstrated. Our proposed forecasting framework used in this paper employs GP models by incorporating three additional features fundamental to our modeling process. The three features are: 1. Automatic Relevance Determination (ARD), used to select model data points for inputs; 2. Predicted stock prices from the NMP model are corrected before any volatility forecasting. Lastly, detailed adjustments were applied to improve our forecasting accuracy using some adjustments in detail, such as using historical data and a separate model building for high stock prices. The NMP data usually includes market variables such as trading volumes, stock returns, interest rates, and inflation. It is clear that stock returns mainly depend on the actual stock prices. However, we need to find out if any other market variables also play an essential role, and even if we know, we may fail to understand how much the variable can affect the returns. To cater to this case, an ARD is used to investigate the selection of input variables. Two main ways can be used to obtain stock returns from NMP data: 1. by learning directly the model between NMP data and stock returns data using a censored GP and correcting the error in NMP stock price prediction and then building a second model for the relationship between stock prices and derived stock returns. 2. Can be obtained through a belief, underpinned by a large body of empirical analysis, that some systematic and stochastic biases are present in the original NMP forecasts. For formality's sake, we denote the first way of modeling stock returns as GP-direct and the second as GP-CPrice (meaning based on corrected price):

For interest, we give a simple schematic diagram of the modeling process, as shown in Figure 1 below.

**Figure1. Model building structure**

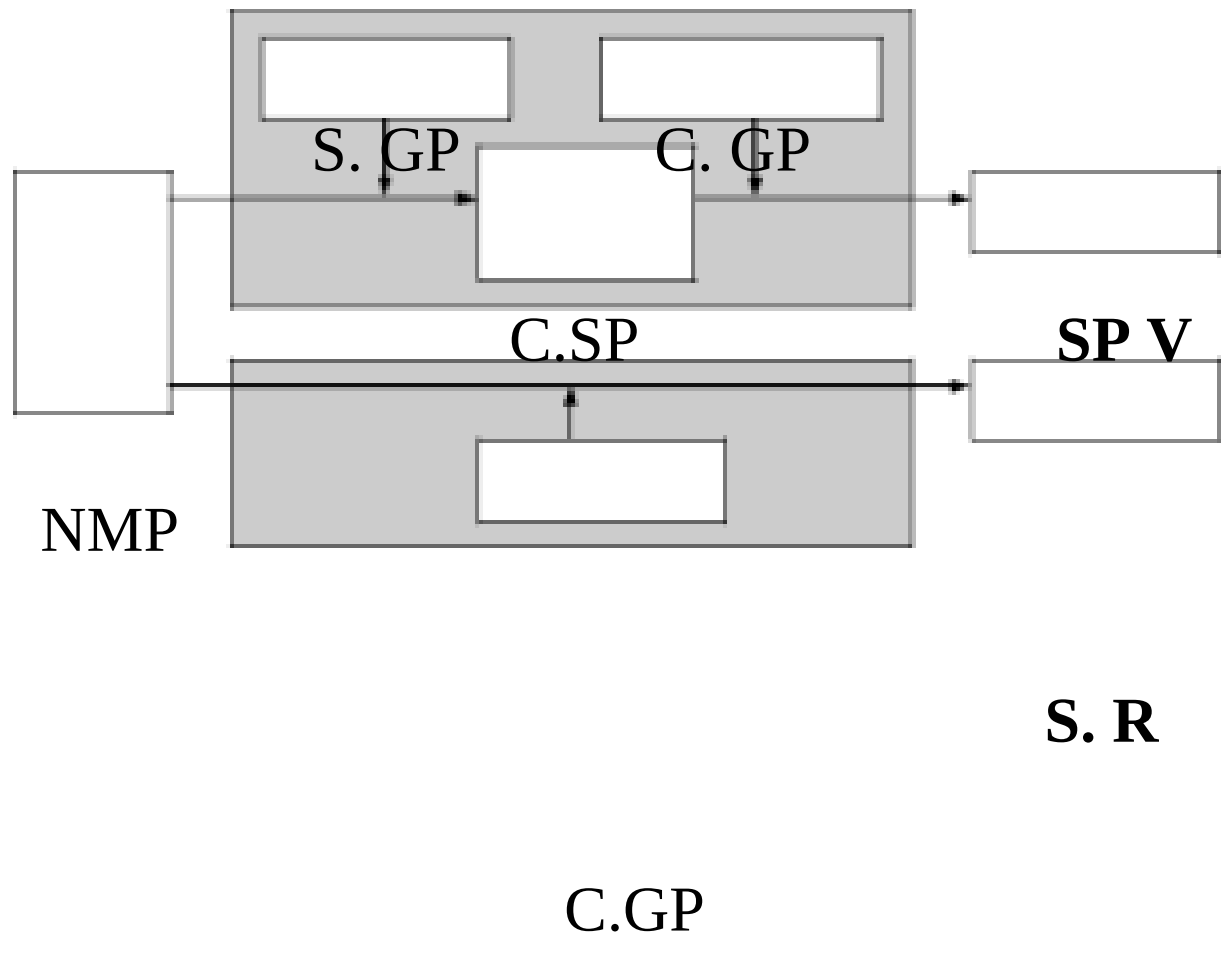

The terms in the diagram above are defined as follows:

NMP- Numerical market prediction

S.GP- Standard Gaussian Process

C.GP- Censored Gaussian Process

C.SP- Censored Stock price

S. PV- Stock price volatility

SR- Stock returns

Further, we then apply our proposed correction model process, considering specific constraints to improve modeling accuracy. First, the method provides forecasts of price volatility and returns. The main aim is to explore the effect of price volatility on stock returns. As mentioned earlier, the primary call of this paper is to develop an efficient price volatility model that can be used to make relevant and frequent volatility estimates. The knowledge of such forecasts and explorations is sound when modeling stock returns, which is the reason behind all investment trades within stock markets.

## 5. Forecasting accuracy evaluation

Evaluating forecasting accuracy and efficiency can be done using several criteria. This study employed two to evaluate our proposed approach and for model evaluation and model comparison: The Root Mean Square Error (RMSE) and the Mean Absolute Error (MAE). We defined the error measures as follows:

$$e_t = y_t - \hat{y}_t \ldots\ldots\ldots..(4)$$

$$RMSE = \left(\frac{1}{n}\sum_{i=1}^{n} e_i^2\right)^{\frac{1}{2}} = \sqrt{\frac{1}{n}\sum_{i=1}^{n} e_i^2} \ \ldots\ldots.(5)$$

$$MAE = \frac{1}{n}\sum_{i=1}^{n}|e_i| \ldots\ldots\ldots\ldots\ldots\ldots(6)$$

Here, $y_t$ denotes the actual observation value at time$t$, $\hat{y}_t$ represents the forecast value for the same period, $n$ is the number of forecasts, and the error is denoted by $e_i$. The forecasting error threshold for the above-specified methods is 10%. The accuracy should be less or equal to 10%.

### 5.1. Experiments

**Stock market price-return charts**

We present three main charts for price evolution, volatility, and return pattern. The idea is to visualize and identify the market behavior of the JSE stock index over the time horizon considered. Additionally, the volatility chart corresponds to the forecasted short-term market volatility.

**Figure 2a): Stock Price evolution**

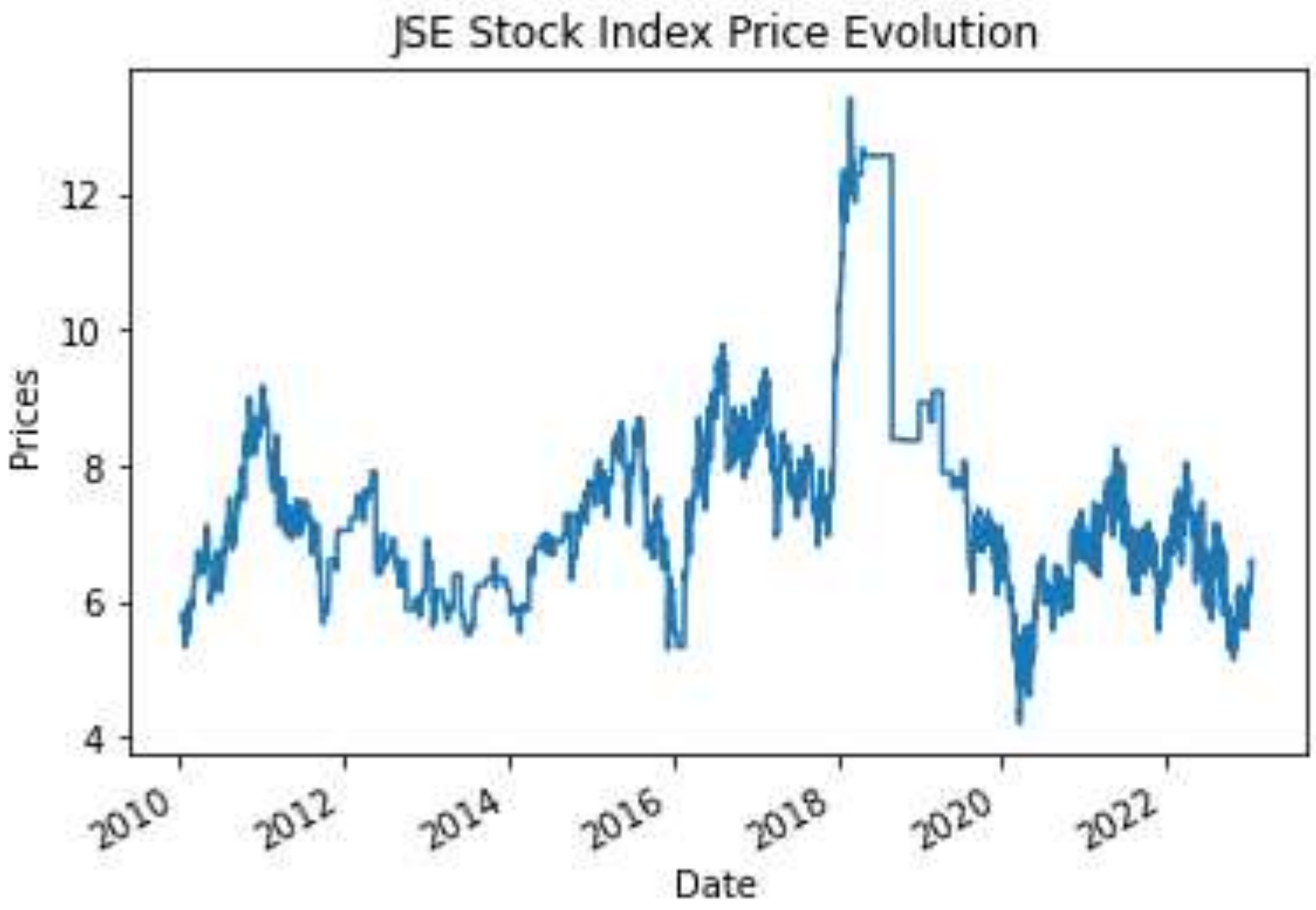

**Figure 2b): Stock price volatility**

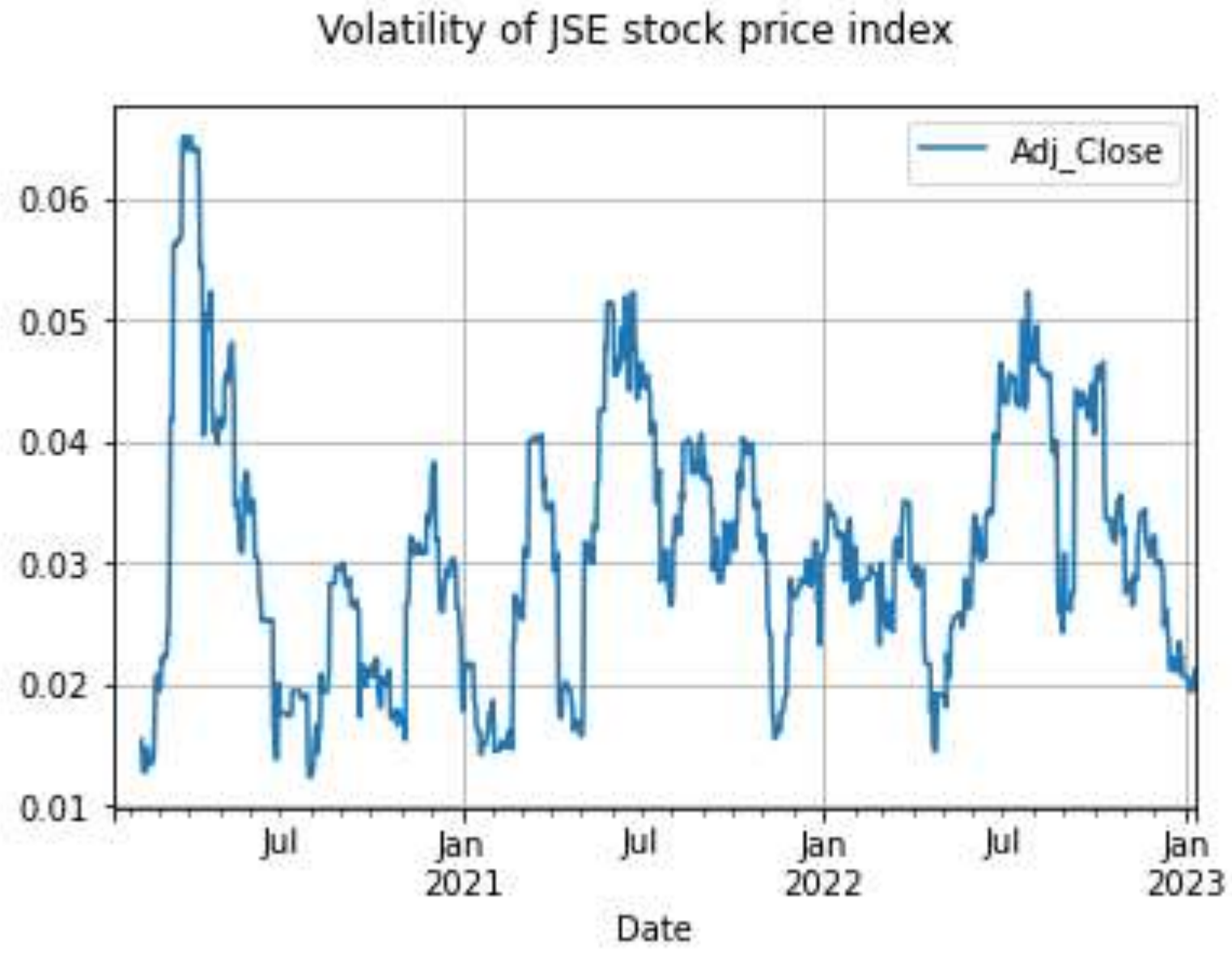

Figure 2b denotes the JSE stock market's short-term price volatility with an interest in identifying the safe and risky regions for investment.

**Figure 2c): Short-Term Stock Returns**

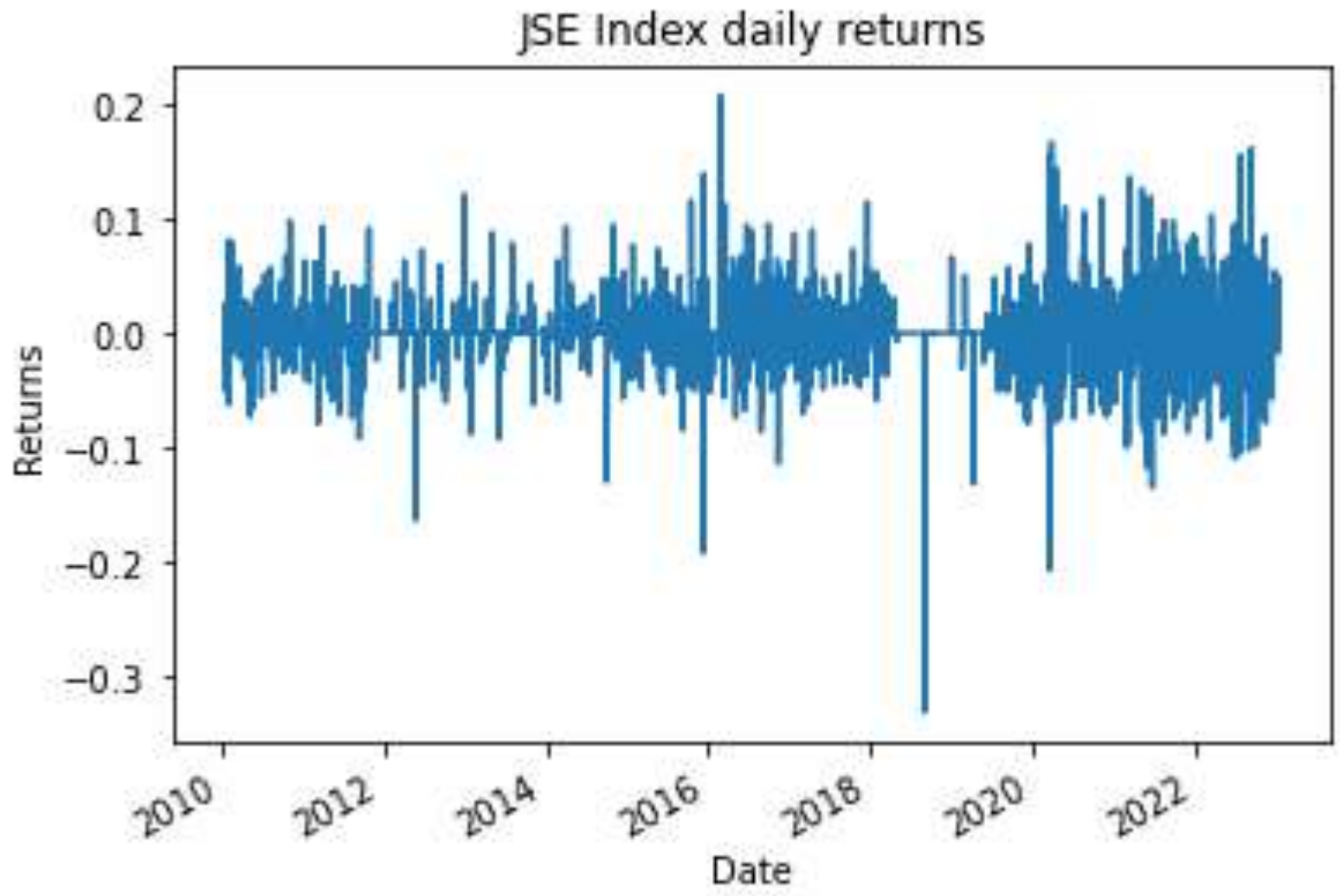

Figure 2c above shows the market behavior of stock returns from 2010 to 2023, with a sharp jump in 2019. The jump comes from the COVID-19 pandemic shock, and thus, our results are ideal and informative to interested but risk-averse investors. Investors should consider all possible external forces when making short-term investments, just like in long-term ones.

***Additional Comments:*** This paper uses two datasets based on JSE records and estimates to evaluate our approach. We first compared the implied price volatility with the forecasted price volatility. We used the Root Mean Square Error (RMSE) to compute the forecasting error to validate our modeling approach. Secondly, we used the Mean Absolute Error (MAE) method to validate our stock return forecasts, where we compared the actual returns and the estimated returns. The two sets of pairwise data are independent as they are extracted differently.

**Table 1: Implied volatility versus estimated volatility**

| Time (month) | Implied volatility (%) (1) | Estimated volatility (%) (2) | Errors (3) |
|---|---|---|---|
| 1 | 0.280 | 0.279 | 0.001 |
| *2* | 0.23 | 0.223 | 0.007 |
| 3 | 0.296 | 0.283 | 0.013 |
| *4* | 0.178 | 0.170 | 0.008 |
| *5* | 0.312 | 0.291 | 0.021 |
| *6* | 0.337 | 0.32 | 0.017 |
| *7* | 0.117 | 0.112 | 0.005 |
| *8* | 0.49 | 0.48 | 0.01 |
| *9* | 0.413 | 0.419 | (0.006)** |
| *10* | 0.60 | 0.60 | 0.000 |
| 11 | 0.556 | 0.532 | 0.024** |
| 12 | 0.80 | 0.798 | 0.003 |

***RMSE=0.012168****

*Root Mean Square Error (RMSE) results*

As shown in Table 1, the implied volatility coefficients are not significantly different from the estimated coefficients. As implied volatility measures the realized volatility associated with price changes (short and long-term), our estimated volatilities proved more reliable. Small RMSE values will indicate this. Another exciting outcome is that volatility is high during holidays and weekends due to less market stability and reduced liquidity levels.

**Table 2: Actual stock returns versus forecasted stock returns**

| Time (month) | Actual returns (%) (1) | Forecasted Returns (%) (2) | Errors (3) |
|---|---|---|---|
| 1 | 0.360 | 0.379 | (0.019)** |
| *2* | 0.655 | 0.635 | 0.02 |
| 3 | 0.698 | 0.6901 | 0.0079 |
| *4* | 0.738 | 0.738 | 0.000** |
| *5* | 0.712 | 0.691 | 0.021 |
| *6* | 0.831 | 0.832 | (0.001)* |
| *7* | 0.8273 | 0.8121 | 0.0152 |
| *8* | 0.749 | 0.748 | 0.001 |
| *9* | 0.713 | 0.409 | 0.304** |
| *10* | 0.635 | 0.62 | 0.015 |
| 11 | 0.756 | 0.732 | 0.024 |
| 12 | 0.57 | 0.568 | 0.002 |

*MAE=0.03584=3.58%*

*Mean absolute error (MAE) results:* As depicted in the above-presented table. Our model proved more accurate, as indicated by small MAE values. The forecast errors are insignificant, indicating slight deviations of our estimated returns from the actual (observed returns).

### 5.2. Model evaluation

As a preliminary step, the ARD was applied to determine which NMP variables should be included as inputs to the correction model. For clarity, we tabulated the measured conditional stock price values as our target variable in the presence of other selected variables-trading volumes, insider news, inflation, exchange rates, and stock returns.

**Table 3: ARD results for stock prices at JSE**

| Variable | Modelling period (1) | Modelling period (2) |
|---|---|---|
| *Stock prices* | 0.380 | 0.313 |
| *Trading volumes* | 0.33 | 0.364 |
| *Trade Frequency* | 0.206 | 0.21 |
| *Interest rates* | 0.10 | 0.08 |
| *Inflation* | 0.08 | 0.121 |
| *Insider news* | 0.27 | 0.29 |

*Variable effect in percentages: Higher percentage, higher effect.*

The intensity and effect of the variables on our prediction accuracy for stock returns and volatility are all different. However, we noted that stock prices and trading volumes impact our prediction much more than the other factors. At the same time, volatility is mainly influenced by trading volumes and insider news in the market. Therefore, we use trading volumes and stock prices (historical) to predict our stock returns and associated short-term volatility as inputs in the GP correction process.

**Simulation Results**

This section presents the results of our stock return-prediction framework and price volatility against some benchmarks. We employed the persistence and multi-layer perceptron (MLP) neural network models. We used the approach by [29] and [30], where they applied the MLP method to forecast wind power generation. The idea behind the persistence method is that it simply uses the current value as the forecast, which means that at the time, $t$, the prediction, $\hat{y}_{t+1} = \hat{y}_{t+2} = \cdots \hat{y}_{t+30} = y_t$.

Since stock data prices are Markovian, we excluded the historical data in this model at the pre-processing stage), The Markov property states that the current/present values better explain the future values of a stock or its price than its past. As such, we use the current stock price and volume data in this model and calculate both the stock returns and price volatilities by stock price-yield curve and volatility smile functions, respectively, which can be obtained by training the historical dataset. MLP networks are seen in application to short-term wind power forecasting than to stock markets data, for example [30]. This study is making use of the networks. An MLP-based model that corrects stock prices and predicts stock returns is chosen for comparison. Using the empirical results (model comparison on a validation set), the first MLP model, which corrects stock prices, used NMP stock prices, trading volumes, and exchange rates as input variables and measured returns as the output variable, with an 11-neuron hidden layer. The second part of the MLP-Stock price model uses corrected stock prices as input and has an 8-neuron hidden layer, then outputs the final prediction of stock returns. Our empirical results conclude that one well-trained forecast model can be applied to other financial asset data of the same type at JSE. The results of using the proposed model to the test datasets are shown in Tables 4 and 5

**Table4. Stock returns forecast error**

| **Model** | RMSE (%) | MAE (%) | NMAPE (%) |
|---|---|---|---|
| CReturns | 14.59 | 12.79 | 10.45 |
| GP-Direct | 12.40 | 12.09 | 9.53 |
| GP CReturns | 8.36 | 7.29 | 5.73 |

**Table5. Price volatility forecasts error**

| **Model** | RMSE (%) | MAE (%) | NMAPE (%) |
|---|---|---|---|
| MLP-CPrice | 15.59 | 12.79 | 9.26 |
| GP-Direct | 13.10 | 11.49 | 7.53 |
| GP-CPrice | 9.36 | 8.69 | 4.73 |

**Combined comments**

From Tables 4 and 5, we note that the proposed GP-Stock Price model performs better than the other models, especially presenting an outstanding performance in the 1-3 hours forecast horizon. In terms of MAE, an accuracy improvement of 17.98% is required. If compared to the MLP-Stock Price model, the gain would be 11.61%. The normalized mean absolute percentage error (NMAPE) is the best measure of the forecasting error in our study. This is supported by its ability to provide non-deviating estimates. For easy reference, the NMAPE is calculated as follows:

$\frac{1}{n}\sum_{i=1}^{n}\left|\frac{e_i}{M}\right| \times 100$, where $n$ is the number of sample items, $and\ M$ is the market type. In this case, we have the stock market (Johannesburg stock exchange.

## 6. Conclusions

Short-term volatility forecasting for stock prices and returns is an essential but challenging task, considering the uncontrollable and stochastic nature of prices and returns. This paper investigated a combination of numeric and probabilistic models: A Gaussian Process (GP) combined with a Numerical market prediction (NMP) model was applied to one-day-ahead return forecasting. Specific methods were employed to improve the forecast accuracy: predicted stock prices are corrected by GP before it is used to forecast stock returns. A censored GP is applied to build the price-return model, mainly to cater for unobserved or missing price records; ARD is used to choose influential NMP variables as inputs to each model; for very short-term forecasts, historical data is added into the modeling process; and a high stock prices subset is treated separately by building a single forecast model as we considered it as a particular case. The simulation results show that, compared to an MLP-Stock Price model, the proposed model has around 11% improvement in forecasting accuracy. Hence, the effectiveness and performance of the GP-Stock Price model are proved. We proved that the GP performs better than other time series stochastic models such as GARCH (1,1) and ARIMA models widely used in volatility forecasting. Therefore, this paper suggests future works to be carried out on high-frequency trades using the proposed model to make informative forecasts on short-term volatilities.